\documentstyle[aps]{revtex}

\def\nap{{\nabla_{\!\!\!{\tiny \perp}}}}
\def\D{\Delta}
\def\sz{S^{(0)}}
\def\so{S^{(1)}}
\def\st{S^{(2)}}
\def\ho{{h^{(1)}}}
\def\htwo{{h^{(2)}}}

\def\iin{\int_{0}^{\infty}}

\def\izl{\int_{0}^{\lambda}}

\def\iorec{\int_{\eta_o}^{\eta_{\rm rec}}}
\def\iol{\int_{\eta_o}^{\lambda}}

\def\jo{j_{1}}
\def\jw{j_{2}}

\def\sol{\so(x(\la),\eta(\la))}
\def\hol{\ho_{ab}(x(\la),\eta(\la))}

\def\deleta{{ \partial\over \partial\eta}}
\def\delsqeta{{ \partial^2\over \partial\eta^2}}

\def\lsb{\left [}
\def\rsb{\right ]}
\def\la{\lambda}
\def\lap{\lambda^{'}}

\input epsf
\def\plotone#1{\centering \leavevmode
\epsfxsize= 0.8\columnwidth \epsfbox{#1}}

\begin{document}
\draft
\preprint{CITA-97-25}
\title { Skewness in the Cosmic
Microwave Background Anisotropy \\ from Inflationary Gravity Wave
Background} 
\author{Somnath Bharadwaj} 
\address{ Mehta Research Institute,10 Kasturba Gandhi Marg, Allahabad 211 002,
India.\\ email: somnath@mri.ernet.in}
\author{Dipak Munshi}
\address
{Queen Mary and Westfield College, London E1 4NS, United Kingdom
\\email: D.Munshi@qmw.ac.uk}
\author{Tarun Souradeep} 
\address{ Canadian Institute for Theoretical Astrophysics,
University of Toronto, Toronto, ON M5S 3H8, Canada\\ email: tarun@cita.utoronto.ca} 
\date{\today}
\maketitle
\begin{abstract}
In the context of inflationary scenarios, the observed large angle
anisotropy of the Cosmic Microwave Background (CMB) temperature is
believed to probe the primordial metric perturbations from inflation.
Although the perturbations from inflation are expected to be gaussian
random fields, there remains the possibility that nonlinear processes
at later epochs induce ``secondary'' non-gaussian features in the
corresponding CMB anisotropy maps. The non-gaussianity induced by
nonlinear gravitational instability of scalar (density) perturbations
has been investigated in existing literature. In this paper, we
highlight another source of non-gaussianity arising out of higher
order scattering of CMB photons off the metric perturbations. We
provide a simple and elegant formalism for deriving the CMB
temperature fluctuations arising due to the Sachs-Wolfe effect beyond
the linear order. In particular, we derive the expression for the
second order CMB temperature fluctuations. The multiple scattering
effect pointed out in this paper leads to the possibility that tensor
metric perturbation, i.e., gravity waves (GW) which do not exhibit
gravitational instability can still contribute to the skewness in the
CMB anisotropy maps. We find that in a flat $\Omega =1$ universe, the
skewness in CMB contributed by gravity waves via multiple scattering
effect is comparable to that from the gravitational instability of
scalar perturbations for equal contribution of the gravity waves and
scalar perturbations to the total {\it rms} CMB anisotropy. The
secondary skewness is found to be smaller than the cosmic variance
leading to the conclusion that inflationary scenarios do predict that
the observed CMB anisotropy should be statistically consistent with a
gaussian random distribution.
\end{abstract}

\section{Introduction}

Since its discovery by Penzias and Wilson \cite{pen_wil67}, the Cosmic
Microwave Background (CMB) has proved to be an extremely significant
observational guide in our quest towards understanding the universe.
The detection of tiny anisotropies in the CMB by the COBE - DMR group
\cite{smoot92} was an important milestone in the study of the universe
and the understanding the large structure that we see around us. The
COBE detection has opened up a fresh avenue of investigation and has
been followed by a host of new developments both on the observational
and theoretical fronts~\cite{bon96}.

The idea of incorporating an inflationary phase in the early universe
\cite{infl} has gained wide acceptance in the last decade and is
perhaps the most prevalent scenario within which one attempts to
understand the universe. It was realised soon after the notion of an
inflationary scenario was put forward that besides resolving some long
standing problems of the Big-Bang model of cosmology, inflationary
models also predict the form of the power spectrum of the primordial
scalar metric fluctuations which could seed the formation of the large
scale structures observed in the present universe\cite{infl_dp}. In
fact, both gravity waves (GW), i.e., tensor metric fluctuations
\cite{infl_gw}, as well as adiabatic density perturbations (related to
the scalar metric fluctuations) arise as natural consequences of the
inflationary scenario, due to the superadiabatic amplification of
zero-point quantum fluctuations occurring during inflation.  As gravity
waves \cite{infl_gw_cmb} and scalar density perturbations enter the
horizon, they induce distortions in the cosmic microwave background
(CMB) through the Sachs-Wolfe effect\cite{sac_wol67}.  The relative
contribution of the gravity waves and the adiabatic perturbations is
linked to the specific model of inflation \cite{infl_qsqt}.  The
spectral index of the power spectrum of initial perturbations can be
inferred directly from the CMB anisotropy measurements.

The measurements of the power spectrum of CMB temperature fluctuations
have till date been found to be consistent with inflationary scenarios
of the early universe. In particular, the spectral index inferred from
the COBE - 4year data (see for eg.\cite{ben96}) is consistent with the
near scale-invariant spectrum of fluctuations generically predicted by
inflationary scenarios. Another generic prediction of inflation is
that the metric perturbations generated are gaussian random fields. At
the linear order, the CMB anisotropy produced would reflect the
gaussian nature of initial perturbations.  However, one cannot apriori
deny the possibility that non-linear corrections to the growth of
perturbations and to the Sachs-Wolfe effect could induce some
non-gaussian features in the CMB even for gaussian initial metric
perturbations.

Non-zero skewness is a definite signature of non-gaussianity in a
distribution. For gaussian initial perturbations the skewness in the
CMB appears only beyond the linear approximation. At the leading order
the skewness arises from two distinct effects.  The {\em first effect}
is that initially gaussian metric perturbations become non-gaussian
when the lowest order nonlinearity becomes important in the course of
their evolution due to gravitational instability. The non-linear
component of the metric perturbations are non-gaussian and introduce
non-gaussian anisotropies in the CMB through a linear Sachs-Wolfe
relation at the corresponding order.  The {\em second effect} is that
the gaussian metric perturbations introduce non-gaussian anisotropies
in the CMB due to  second order (double) scattering of the photon off
the linear order metric perturbations. This arises from the second
order terms in the Sachs-Wolfe relation. All previous discussions of
skewness in the CMB, have been limited to the the estimation of only
the first effect, i.e., nonlinearity (and consequent non-gaussianity)
due to gravitational instability
\cite{luo_sch93,mun_sour95,gang95}. The tensor component of metric
perturbations (GW) does not exhibit gravitational instability,
consequently the possibility of non-gaussianity in the CMB caused by
the GW background has been entirely ignored in previous literature.

In this paper, as an illustration of the second effect, we calculate
the CMB skewness produced by a gaussian stochastic linear gravity wave
background generated during inflation. In the context of $\Omega =1$,
flat FRW models, the magnitude of the effect considered here appears
to be comparable to the corresponding estimate of CMB skewness arising
from the gravitational instability of scalar metric perturbations
\cite{mun_sour95}.

In $\S$\ref{sectwo}, we outline the basic formalism involved in
estimating the skewness in the CMB and present a very general approach
for obtaining higher order corrections to the Sachs-Wolfe effect for a
general cosmological perturbation.  In the following section
($\S$\ref{secthree}) we estimate the skewness in the CMB anisotropy
that would arise from a inflationary gravity wave background for a
range of values of the spectral index.

\section{Formalism}
\label{sectwo}

In this section, we outline the basic approach and present the
derivations of results used in our calculation. The first part of the
section contains a brief discussion of the perturbative approach used
in estimating non-gaussianity in the CMB anisotropy. The second part
gives a compact derivation of the CMB anisotropy arising through the
Sachs-Wolfe effect upto second order in the primordial metric
fluctuations.

\subsection { Non-gaussianity and Nonlinearity in the CMB anisotropy}
\label{sectwoA}

It is possible to address non-linear effects in the CMB within a
perturbative framework by expanding the temperature fluctuations, $\D
T/T$, in orders corresponding to the powers of the initial metric
perturbation as~:

\begin{equation}
\frac{\D T}{T} = \left(\frac{\D T}{T}\right)^{(1)} + \left(\frac{\D
 T}{T}\right)^{(2)} + \left(\frac{\D T}{T}\right)^{(3)} +\dots ~~~.
\label{cmb_nl_exp}
\end{equation}

\noindent 
Given that the {\em initial} metric perturbations from inflation are
{\em linear} and {\em gaussian}, any {\em non-gaussian feature} in the
CMB maps can only arise from the {\em higher order temperature
anisotropy} such as $(\D T/T)^{(2)}$. We shall call this higher order
effect --- {\em secondary non-gaussianity}. (The term ``secondary" is
used to denote the effects which take place after recombination. The
effects prior to recombination are ``primary".)

A non-vanishing skewness is a definite signature of non-gaussianity in
a distribution. At the linear order, the mean CMB skewness
$C_3^{(3)}(0)=\langle (\Delta T^{(1)}/T)^3 \rangle =0$ where $\langle
\rangle$ denotes averaging with respect to different realizations of
stochastic space-time metric perturbations of the FRW cosmological
model which produce $\Delta T/T$.  Substituting the expansion
(\ref{cmb_nl_exp}) into the expression $C_3(0) = \langle (\Delta
T/T)^3 \rangle$, it is clear that the leading order (in powers of the
initial linear metric perturbations) contribution is at the fourth
order,

\begin{equation}
C_3^{(4)}(0)=3\bigg\langle \left({\Delta T\over T}\right)^{(1)}
\left({\Delta T\over T}\right)^{(1)}
\left({\Delta T\over T}\right)^{(2)}~\bigg\rangle~. 
\label{skw_four}
\end{equation}

\noindent (In this paper, we deal only with $C_3^{(4)}(0)$, and hence
the superscript denoting the order has been dropped in the rest of the
text).  It is clear from the above expression that the mean skewness
depends not only on the magnitude of $\Delta T^{(2)}/T$ but also on
the extent of correlation of this term with the linear order terms,
$\Delta T^{(1)}/T$.  For example, in the case of scalar perturbations
the second order terms which arise due to the non-linear evolution of
density perturbations  grows linearly with the expansion of the
universe and can attain values $\Delta T^{(2)}/T \approx 0.1~ \Delta 
T^{(1)}/T$ at late times \cite{mar_san92}. However, in a flat $\Omega
= 1$ universe, the linear order term contributes only close to the
surface of recombination ($\eta \approx\eta_{\rm rec}$) and the second
order term attains its largest value only at late times
($\eta\approx\eta_0$). Consequently in the final result for the mean
skewness, the decay of correlation between the linear and the second
order term over this large physical separation ($\approx
\eta_o$) along the line of sight attenuates the effect
of the growth of the second order term, leading to a very modest value
for $C_3(0)$ \cite{mun_sour95}.  It was also pointed out in the same
paper that the mean skewness which arises due to  weakly
non-linear density perturbations is expected to be somewhat larger in
models where the linear gravitational potential changes at late times leading
to a significant linear order integrated Sachs-Wolfe contribution at
late times (eg. CDM+$\Lambda$, $\Omega \neq 1$ models).

Even in a flat $\Omega = 1$ universe, contribution to linear order
integrated Sachs-Wolfe effect comes from inflationary tensor
perturbations (gravity waves).  Consequently, one expects that the
correlation between the linear and second order terms is not
attenuated in this case leading to larger values of $C_3(0)$. The
second order $\Delta T^{(2)}/T$ in the case of gravity waves comes
only from double scattering since GW do not exhibit any gravitational
instability. Scalar perturbations also give rise to second order
anisotropy through double scattering. However, for flat, $\Omega = 1$
models the contribution to the mean skewness is expected to be even
smaller than that from gravitational instability considered in
\cite{mun_sour95}. This can be seen from the fact that $\Delta
T^{(2)}/T$ from double scattering too has contributions only at late
times implying attenuated correlation with the linear term contribution
close to the surface of recombination.

\subsection { Second order CMB anisotropy from the Sachs-Wolfe effect}
\label{sectwoB}

In a perfectly isotropic universe the CMB would have the same
temperature in all directions on the celestial sphere. If, however,
the cosmological metric is perturbed, the temperature observed today
fluctuates over the celestial sphere.

The dominant contribution at large angular scales ($\theta > 1^\circ$)
to the observed temperature fluctuations comes from the change in the
frequency of any CMB photon as it travels from the surface of last
scattering to us \footnote{ We assume instantaneous recombination
which is an excellent approximation to standard recombination for
calculating the CMB anisotropy at large angular scales ($\theta >
1^\circ$).}.  In the case of an isotropic universe, the overall
increase in the scale factor $a(\eta_0)/a(\eta_{\rm rec})$ redshifts
the entire Planckian distribution of photons leading to a Planckian
distribution at a lower temperature given $T_{\rm rec}/T_0 =
a(\eta_0)/a(\eta_{\rm rec}) $. The presence of the perturbations
$h_{ab}(\eta, {\bf x})$, produces an additional change in the
frequency and direction (momentum) of a photon as it moves in and out
of the fluctuating metric perturbations.

 We consider the trajectory of a photon (or ray) in a perturbed flat
FRW universe and work in a synchronous coordinate system where the
line element is of the form

\begin{equation}
ds^2= a(\eta)^2\lsb - d\eta^2 + \left( \delta_{ab} + h_{ab}\left(\eta, 
x\right)\right)~ dx^a dx^b\rsb\, .
\label{pmetric}
\end{equation}
In the above, $h_{ab}=h^{(1)}_{ab} (\sim \epsilon) + h^{(2)}_{ab} (\sim
\epsilon^2$) is the metric perturbation, and $\epsilon \ll 1$ is a
small number characterizing the amplitude of deviations from the
unperturbed background FRW universe.

The photon trajectory can be obtained by perturbatively solving the
eikonal equation for the phase $S(x,\eta)$. The eikonal equation
for a photon propagating in a spacetime with metric, $g_{\mu\nu}$, is

\begin{equation}
\frac{\partial S}{\partial x^{\mu}} \frac{\partial S}{\partial
x^{\nu}}~ g^{ \mu \nu}=0 \, ,
\end{equation}
(analogous to the Hamilton-Jacobi equation for a massive
particle). The frequency and the direction of the photon can be
obtained from the phase $S(x,\eta)$ using
\begin{equation}
\omega(x,\eta)=-\frac{1}{a(\eta)} \deleta S(x,\eta),\hspace{1cm}
k_a(x,\eta)=\nabla_a S(x,\eta) \,. \label{eq:a3}
\end{equation}

Retaining terms to order $\epsilon^{2}$ when inverting the metric
given by the line element in equation.(\ref{pmetric}), the eikonal
equation~(\ref{eikoneqn}) becomes 

\begin{equation}
- \left( \deleta S \right)^2 + ( \delta^{ab} -
{\ho}^{ab} -{\htwo}^{ab} + {\ho}^a_c {\ho}^{cb}) \nabla_aS \nabla_b S  = 0 \;.
\label{eikoneqn} 
\end{equation}
where we use the background spatial metric $\delta_{ab}$ to raise and
lower the spatial indices. The phase $S(x,\eta)$ can be expressed in a
perturbative expansion in powers of $\epsilon$, as
\begin{equation}
 S = \sz + \so(\sim \epsilon) + \st(\sim \epsilon^{2})\;.
\end{equation}
and substituting the above in equation (\ref{eikoneqn}) we obtain the 
zeroth order equation 
\begin{equation}
 -\left(\deleta \sz  \right)^2 + \nabla_a \sz \nabla^a\sz = 0\, ,
\end{equation} 
the first order equation 
\begin{equation}
- \deleta \sz \deleta \so + \nabla_a \sz \nabla^a \so =
\frac{1}{2} {\ho}^{ab} \nabla_a \sz \nabla_b \sz\, ,
 \label{eq:a4}
\end{equation}
and the second order equation  
\begin{eqnarray}
- \deleta \sz \deleta \st + \nabla_a \sz \nabla^a \st &=&
 \frac{1}{2} \left(  \deleta \so \right)^2  -
\frac{1}{2}  \nabla_a\so \nabla^a \so  + {\ho}^{ab} \nabla_a \so
\nabla_b \sz \nonumber \\ 
&+& \frac{1}{2} \lsb{\htwo}^{ab} -{\ho}^{ac} {\ho}_c^b \rsb \nabla_a \sz
 \nabla_b \sz \,.\label{eq:a5}
\end{eqnarray}

The solutions to these equations correspond to a family of rays (or
null geodesics) in a perturbed FRW universe.  The solution to the
zeroth order solution is
\begin{equation}
{\sz}(x,\eta) = k_a x^a - \eta + C\;. 
\end{equation}

\begin{figure}
\plotone{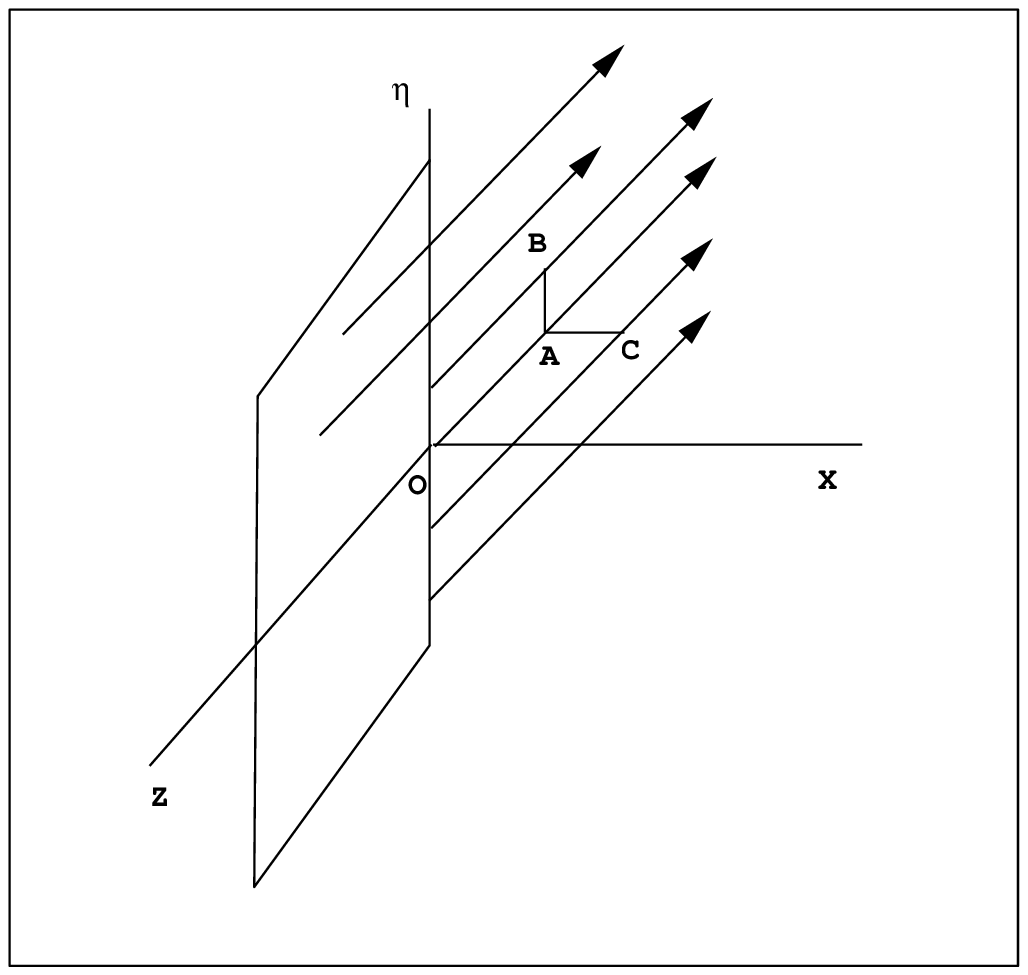}
\caption[]{A sketch of the unperturbed rays of CMB photons propagating
along the $\bf x$ axis is shown (the third spatial dimension is
suppressed).  The figure serves as a guide to the arguments used in
$\S$\ref{sectwoB} to compute the temporal and spatial derivatives of
the phase $S(\eta,{\bf x})$. The figure clearly illustrates the
distinction between spatial derivative taken along the direction of
photon propagation and spatial derivative taken perpendicular to the
direction of photon propagation.}
\label{fig}
\end{figure}

This corresponds to a family of trajectories for the photons with
frequency $1/a(\eta)$ moving in the ${\bf k}$ direction. In figure
\ref{fig}, we show some of these trajectories for the case where ${\bf
k}$ is along the $\bf x$ axis (the $\bf y$ direction has not been
shown). The observer O sits at the origin of the spatial coordinate
system and measures the frequency of the photons at the present epoch
$\eta_0$. Using $\lambda$ as a parameter along the rays, this family
of rays can be expressed as
\begin{equation}
{\bf x}(\la)={\bf k} \la  + {\bf X}
\end{equation}
and 
\begin{equation}
\eta(\la)=\la + \eta_{in}
\end{equation}
with the condition that ${\bf X}$ is perpendicular to ${\bf k}$. 
Here different values of $(\eta_{in},{\bf X})$ correspond to different 
rays and $(\eta_{0},0)$ corresponds to  the ray that the observer O 
sees at present traveling in  the ${\bf k}$ direction .

Substituting the first order solution in equation (\ref{eq:a4}), we
rewrite the LHS as a derivative along the zeroth order
trajectory, i.e., with respect to the parameter $\lambda$ as
\begin{equation}
\frac{d}{d \la} \sol = \frac{1}{2} \hol k^a k^b \label{eq:xx}
\end{equation}
which can be easily integrated to obtain the first order solution,
\begin{equation}
\so(\eta(\la_1),x(\la_1)) = {1 \over 2} \int_0^{\la_1} h^{(1)}_{ab}
(\eta(\la),x(\la)) k^a k^b d \la  \;. \label{eq:a7}
\end{equation}

The frequency of the photon is given by the time derivative of $\so$
given by equation (\ref{eq:a7}).  At the point A (refer to figure~\ref{fig})
the time  derivative is the $\Delta \eta \to 0$ limit of 
\begin{equation}
\frac{S(B)-S(A)}{\Delta \eta} \label{eq:ab8}
\end{equation}
where $\Delta \eta$ is the difference in $\eta$ between the points B and A.   

The point A lies on the trajectory ${\bf x}={\bf k} \la $, $\eta=\la + \eta_0$ 
and it corresponds to a value $\lambda_1$ for the parameter $\la$. The point B
lies on a different  trajectory  ${\bf x}={\bf k} \la $, $\eta=\la + \eta_0 +
\Delta \eta$ and it corresponds to the same value of the parameter
i.e.,  $\lambda_1$. We then have
\begin{equation}
S(B)-S(A)= {1 \over 2} \int_0^{\la_1} 
\lsb h^{(1)}_{ab} (\eta(\la)+\Delta \eta,x(\la)) -
h^{(1)}_{ab} (\eta(\la),x(\la))\rsb k^a k^b d \la 
\end{equation}
which leads to

\begin{equation}
\deleta \so (\eta(\la_1),x(\la_1)) = {1 \over 2} \int_0^{\la_1}
 \deleta h^{(1)}_{ab} (\la) k^a k^b d \la \;.
\end{equation}

Similarly, in the direction parallel to ${\bf k}$ (the $x$-axis in
fig.\ref{fig}), the spatial derivative is given by limit
\begin{equation}
\frac{S(C)-S(A)}{\Delta x} \label{eq:aa8}
\end{equation}
where $\Delta x$ is the difference in $x$ between the points C and A.
C lies on a trajectory ${\bf x}={\bf k} \la $, $\eta=\la + \eta_0 -
\Delta x$ and it corresponds to the value of the parameter $\lambda_1+
\Delta x$. We obtain 
\begin{equation}
S(C)-S(A)= {1 \over 2} \int_0^{\la_1+\Delta x} 
h^{(1)}_{ab} (\eta(\la)-\Delta x,x(\la)) k^a k^b  d \la-
{1 \over 2} \int_0^{\la_1} 
h^{(1)}_{ab} (\eta(\la),x(\la)) k^a k^b d \la\, ,
\end{equation}
and  the spatial derivative in the direction parallel to
${\bf k}$ reads
\begin{equation}
\nabla_c {\so} (\eta(\la_1),x(\la_1)) = {k_c \over 2} \left[  
h^{(1)}_{ab}(x(\la_1), \eta(\la_1))  k^a k^b -\int_0^{\la_1} 
\deleta h^{(1)}_{ab} (\la) k^a k^b d \la \right] \;.
\end{equation}

In the direction perpendicular to ${\bf k}$ we have
 
\begin{equation}
\nabla_c {\so} (x(\la_1), \eta(\la_1)) = {1 \over 2} \int_0^{\la_1}
 \nabla_c h^{(1)}_{ab} (\la) k^a k^b d\la\;.
\end{equation}
Substituting the derivatives of $\so$ in the second order
equation~(\ref{eq:a5}) and using $ {\nap}_a=\nabla_a-k_a k^b \nabla_b$
to denote the spatial derivative in the direction perpendicular to
${\bf k}$ we obtain
\begin{eqnarray}
S^{(2)}(x, \eta) &=&  \int_0^{\la_1} d \la \, \,{\Bigg \{} \frac{3}{8}
 \ho_{ab}(\la)
k^a k^b  \ho_{cd}(\la) k^c k^d  - \frac{1}{4}  \ho_{ab}(\la) k^a k^b
\izl \deleta {\ho}_{ij} (\lap) k^i k^j d \lap \nonumber \\
&+& \frac{1}{2} \lsb {{\htwo}}^{ij}(\la)-{{\ho}}^{il}(\la)
{{\ho}}_l^j(\la) \rsb k_i k_j
+\frac{1}{2}  {\ho}^{ab}(\la) k_a \izl 
\nap_b {\ho}_{ij} (\lap) k^i k^j d \lap  
\nonumber \\
&-&\frac{1}{2} \lsb \left( 
\frac{1}{2} \izl { \nap_a {\ho}_{ij}} (\lap) k^i k^j d \lap   \right)
\left( \frac{1}{2} \izl {\nap^a {\ho}_{lm}} (\lap) k^l k^m d \lap \right) 
\rsb  \;\;\; {\Bigg \}}  \label{eq:a6}.
\end{eqnarray}

The perturbative solution for $S({\bf x}, \eta)$ can now be used to
obtain the frequency of a photon at any point along its
trajectory. Although we are interested in a photon that leaves the
last scattering surface at $\eta_{rec}$ and reaches the observer at
$\eta_0$, it is convenient to consider a photon traveling backwards
in time from the observer to the last scattering surface.  Keeping
terms upto order $\epsilon^2$ we find that the photon that is observed
at the frequency $\omega_o=1/a(\eta_o)$ left the last scattering
surface with the frequency,
\begin{equation}
\omega_e=\frac{1}{a(\eta_e)} \lsb 1-
\frac{\partial}{\partial\eta}{\so}(x_e,\eta_e)
-\frac{\partial}{\partial\eta}{\st}(x_e,\eta_e) \rsb \label{eq:z1}
\,.
\end{equation}
Invert the above relation we find that a photon which left the last
scattering surface with the frequency $\omega_o=1/a(\eta_e)$ will have
frequency
\begin{equation}
\omega_o=\frac{1}{a(\eta_o)} \lsb 1+
\frac{\partial}{\partial\eta}{\so}(x_e,\eta_e)
+\frac{\partial}{\partial\eta}{\st}(x_e,\eta_e) +
(\frac{\partial}{\partial\eta}{\so}(x_e,\eta_e))^2 \rsb \, ,
\label{eq:z2}
\end{equation}
when it reaches the observer. This relates the observed frequency to
the emitted frequency and the metric perturbations.  Using equations
~(\ref{eq:z1}) and (\ref{eq:z2}) we obtain expressions for the fractional
change in the frequency of the observed photon relative to the
frequency that would be observed if the universe were unperturbed.
Since the CMB photons have a Planckian distribution, (frequency
independent) fractional changes in frequency translates to
fluctuations in the temperature characterizing the distribution.  At
the linear order we recover the familiar (linear order) Sachs Wolfe
effect
\begin{equation}
\frac{\Delta T^{(1)}}{T} =
 -{1 \over 2} \int_{\eta_{\rm rec}}^ {\eta_0}{\partial \over \partial \eta} 
 h^{(1)}_{ab} (x (\lambda),
 \eta(\lambda)) k^a k^b d \lambda~.
\label{eq:b4}
\end{equation}

At the second order the expression for fractional change in CMB
temperature reads
\begin{eqnarray}
\frac{\Delta T^{(2)}}{T}  
&=&  \int_{\eta_0}^{\eta_{rec}} d \la \,
 \,{\Bigg \{} \frac{3}{4} \ho_{ab}(\la)
k^a k^b  \deleta \ho_{cd}(\la) k^c k^d  \nonumber \\
&-& \frac{1}{4}  \deleta \ho_{ab}(\la) k^a k^b
\iol \deleta {\ho}_{ij} (\lap) k^i k^j d \lap
- \frac{1}{4}  \ho_{ab}(\la) k^a k^b
\iol \delsqeta {\ho}_{ij} (\lap) k^i k^j d \lap  \nonumber \\
&+& \frac{1}{2} \lsb \deleta {{\htwo}}^{ij}(\la)-2 {{\ho}}^{il}(\la)
\deleta {\ho}_l^j(\la) \rsb k_i k_j 
 \nonumber\\
&+&\frac{1}{2}\lsb  \deleta {\ho}^{ab}(\la) k_a \iol 
\nap_b {\ho}_{ij} (\lap) k^i k^j d \lap  
+  {\ho}^{ab}(\la) k_a \iol 
\deleta \nap_a {\ho}_{ij} (\lap) k^i k^j d \lap \rsb  
\nonumber \\
&-&\lsb \left( 
\frac{1}{2} \iol {\deleta \nap_a {\ho}_{ij}} (\lap) k^i k^j d \lap   \right)
\left( \frac{1}{2} \iol {\nap^b {\ho}_{lm},} (\lap) k^l k^m d \lap \right) 
\rsb 
 \;\;\; {\Bigg \}}
\nonumber \\
&+&{\Bigg [} {1 \over 2} \int_{\eta_{rec}}^{\eta_0} 
 \deleta{\ho}_{ij} (\la) k^i k^j d \la {\Bigg ]}^2. \label{eq:a8}
\end{eqnarray}
The expression for ${\Delta T^{(2)}}/{T}$ has been independently
obtained by other methods \cite{pyn_car96,hu_scot94}.

The ${\Delta T}/{T}$ calculated here is the fractional increment
(or decrement) with respect to the CMB temperature that would be
observed in a homogeneous and isotropic universe. In practice an
observer would measure the fluctuations in the CMB temperature from
different directions in the sky and in order to relate our calculation
with what is observed we have to remove any monopole component in the
angular distribution of the calculated ${\Delta T}/{T}$.  In
addition, the dipole component of ${\Delta T}/{T}$ is usually
interpreted as arising from the observers peculiar motion and this is
removed before relating the observed ${\Delta T}/{T}$ to the
primordial fluctuations. 

In most previous work on second order CMB anisotropy, only the term
depending on $h^{(2)}$ arising from nonlinear gravitational
instability of scalar density perturbations in the above expression
for ${\Delta T^{(2)}}/{T}$ has been considered. The nongaussianity
from the $h^{(2)}$ term in ${\Delta T^{(2)}}/{T}$ for $\Omega=1$
models has been considered in reference \cite{mun_sour95,gang95}. The
$h^{(2)}$ term in ${\Delta T^{(2)}}/{T}$ for $\Omega<1$ models has
been calculated in reference \cite{san96} to calculate the fourth
order corrections to the power spectrum of CMB anisotropies.

\section{Skewness from Gravity Wave}
\label{secthree}
Gravitational waves can be decomposed into two polarization states
denoted by $+$ and $\times$, and the metric perturbation corresponding
to a gravitational wave with wavelength $\frac{2 \pi}{q}$ traveling
in the $z$ direction can be written as
\begin{equation}
h_{ab}^{\hat z}(\eta,{\bf x})=e^{i q z} h_q(\eta) 
\lsb a^{+}_{q} e^+_{ab}+a^{\times}_{q}
 e^{\times}_{ab}\rsb~\frac{A_G(q)}{q^{\frac{3}{2}}}
\label{gwform}
\end{equation}
where the quantity $A^2_G(q)/q^3$ is the power spectrum of the GW
perturbations. The temporal evolution of the modes of the gravity
waves in a $\Omega=1$, matter dominated FRW universe is given by
\begin{equation}
 h_q(\eta) = {3 \over q\eta} j_1(q\eta) \;,
\end{equation}
where $j_1$ denotes the spherical Bessel function of order one.  
In eqn. (\ref{gwform}), the $a_q^{\alpha}$s are the complex amplitudes
of the two polarization states (henceforth denoted by the superscript
$\alpha$) and $e^{\alpha}_{ab}$ are the usual basis tensors for the
traceless metric perturbations.

A gravity wave in any arbitrary direction $\hat n$ can be obtained by
rotating the coordinate system (or equivalently the wave traveling in
the $z$ direction) so that the  wave travels in the required direction. 
Denoting the rotation by  $R(\hat n)_{ab}$, we can express a 
gravitational wave traveling in the $\hat n$ direction as   
$h_{ab}^{\hat n}= R_a^c (\hat n) R_b^d (\hat n) h_{ab}^{\hat z}$.

Thus we can write metric perturbations corresponding to an  isotropic
stochastic background GW as

\begin{equation}
\hat h_{ab}(\eta,{\bf x}) = \int{ dq d\Omega_{\hat n} \over (2\pi)^3}
q^{1 \over 2}h_q(\eta) e^{i\bf q \bf x} R_{ac}( \hat n )R_{db}({\hat
n}) [a_{\bf q}^+e_{cd}^{+} + a_{\bf q}^{\times}e_{cd}^{\times}] A_G(q)
\;. \label{eq:b2}
\end{equation}  
In the above,  $a^{\alpha}_{\bf q}$ are independent random 
variables which satisfy the relation 
\begin{equation}
<a^{\alpha}_{\bf q} a^{\beta\hspace{.1cm}*}_{\bf q^{'}}>=
\delta^3(\bf q -\bf q^{'})\delta^{\alpha \beta} \; \;.
\end{equation} 
where the angular bracket denotes an ensemble average over different
realizations of the stochastic GW background.

We substitute the expression (\ref{eq:b2}) for the GW metric
perturbation in equations (\ref{eq:b4}) and (\ref{eq:a8}) to calculate
the temperature fluctuation in the CMB at the first and second orders,
respectively. At the linear order, the $\Delta T/T$ produced by
gravity waves has no monopole or dipole component, but at the second
order there is a monopole component which has to be removed. To keep
the algebra at a manageable level we use the ensemble average of
$\Delta T^{(2)}/T$ instead of the angular average of $\Delta
T^{(2)}/T$ to estimate the monopole component and we subtract this
contribution from the second order temperature fluctuation predicted
by equation (\ref{eq:a8}). We next use this to calculate the skewness.

At the leading order the skewness is given by
\begin{equation}
C_3(0) \equiv \bigg \langle \left( \frac{\Delta T}{T} \right)^3 \bigg
\rangle = 3\bigg \langle \left(\frac{\Delta T^{(1)}}{T}\right)^2
\left(\frac{\Delta T^{(2)}}{T}\right)\bigg \rangle~ \;, \label{eq:b3}
\end{equation}
and substituting for expressions for $\Delta T^{(1)}/T$ and $\Delta
T^{(2)}/T$ we obtain
\begin{eqnarray}
C_3(0) &=& \frac{3^5}{2^5 {\pi}^{10}} \iorec d \la_1  \iorec d \la_2 
\iorec d \la_3  \iin  \frac{d q_1}{q_1} \iin \frac{d q_2}{q_2} 
A^2_G(q_1)  A^2_G(q_2) \frac{\jo(q_1 \la_3)}{ q_1 \la_3}
\frac{\jw(q_1 \la_1)}{\la_1} \nonumber \\
& &\frac{\jw(q_2 \la_3)}{\la_3} \frac{\jw(q_2 \la_2)}{\la_2} 
\lsb \left( \frac{3 \pi}{8} \right)^2 \frac{J_3(q_1(\la_3-\la_1))}
{(q_1(\la_3-\la_1))^2} \frac{J_3(q_1(\la_3-\la_2))} {(q_1(\la_3-\la_2))^2} 
-\frac{\jw(q_1(\la_3-\la_1))}{(q_1(\la_3-\la_1))^2}
\frac{\jw(q_1(\la_3-\la_2))}{(q_1(\la_3-\la_2))^2} \rsb \nonumber \\
&-& \frac{3^5}{2^5 {\pi}^{10}} \iorec d \la_1  \iorec d \la_2 
\iorec d \la_3 \int_{\la_3}^{\eta_0}  d \la_4 \iin  \frac{d q_1}{q_1} 
\iin \frac{d q_2}{q_2} A^2_G(q_1)  A^2_G(q_2)
\frac{\jw(q_1(\la_3-\la_1))}{(q_1(\la_3-\la_1))^2}
\frac{\jw(q_2(\la_4-\la_2))}{(q_2(\la_4-\la_2))^2} \nonumber \\
& &\frac{\jw(q_1 \la_1)}{ \la_1} \frac{\jw(q_2 \la_2)}{\la_2}
\lsb \frac{\jw(q_1 \la_3)}{\la_3} \frac{\jw(q_2 \la_4)}{\la_4}
+\frac{\jo(q_1 \la_3)}{q_1 \la_3}\frac{q_2}{\la_4} 
\left( 4 \frac{\jw(q_2 \la_4)}{\la_4} - \jo(q_2 \la_4) \right) \rsb
\nonumber \\
&+& \frac{3^5}{2^5 {\pi}^{10}} \lsb \iorec d \la_1  \iorec d \la_2 
\iin \frac{d q}{q} A^2_G(q) \frac{\jw(q(\la_2-\la_1))}{(q(\la_2-\la_1))^2}  
\frac{\jw(q \la_1)}{\la_1} \frac{\jw(q_1 \la_2)}{\la_2} \rsb^2 + 6 C_2^2(0)\,,
\end{eqnarray}
where $j_n$ and $J_n$ denote spherical and ordinary Bessel functions
of order $n$, respectively. In the case of matter dominated $\Omega_0
=1$ models with three relativistic species of neutrinos, $\eta_{\rm
rec} = \eta_0/49.6$. The final results are however completley
insensitive to the exact value of $\eta_{\rm rec}$. It is interesting
to note that none of the terms in equation (\ref{eq:a8}) which have
spatial derivatives in direction perpendicular to ${\bf k}$ contribute
to the skewness.

We numerically evaluated the above expression to compute the skewness
for different spectral indices of GW power spectrum.  A dimensionless
skewness parameter $S_3$ can be constructed by dividing the skewness,
$C_3(0)$ by the square of the variance, $C_2(0)$. The variance of
$\Delta T/ T$ that arises due to relic gravity waves from inflation is
given by
\begin{eqnarray}
C_2(0)= \frac{3^2}{2^3 {\pi}^5}  \iorec d \la_1  \iorec d \la_2 
\iin \frac{d q}{q} A^2_G(q) \frac{\jw(q(\la_2-\la_1))}{(q(\la_2-\la_1))^2}  
\frac{\jw(q \la_1)}{\la_1} \frac{\jw(q_1 \la_2)}{\la_2}\,.
\end{eqnarray}

\noindent The power spectrum of the initial gravitational wave
perturbation is  assumed to be a scale-free power law, $A^2_G(k) =
A~k^{n_T}$, where $n_T = 0$ corresponds to a scale-invariant
spectrum. Power law models of inflation  produce gravity waves with
$n_T < 0$. We compute the CMB skewness for a broad range in $n_T$
($-0.5\le n_T\le 0$).  We find that the value of $S_3$ varies from 
$ -1.6$ to $-3.1$ for $n_T$ varying from the scale invariant
spectrum to $n_T = -0.5$.  It is interesting to note that the skewness
arising from gravitational instability of scalar perturbations is
comparable for the same range of tilt ( $S_3 \approx  -2.2$ for scale
invariant spectrum \cite{mun_sour95}).

The corresponding observable quantity for the skewness, $C_3(0)$ is
the all sky-average $\bar C_3(0) = (4\pi)^{-1}\int \left(\Delta
T(\theta, \varphi)/T\right)^3d\Omega$ of one particular realization of
the random fluctuations. The value obtained by taking an angular
average over one sky would generally differ from the ensemble average
over all realizations by the (cosmic) variance of the observed
skewness, $\bar C_3(0)$.  The skewness originating due to any effect
would have an observable significance if the predicted signal stands
above the cosmic variance. This is a fundamental limitation and a
minimal requirement. In practice, a detectable signal has to stand
above additional variances such as instrumental noise, finite beam
width of antennas, incomplete sky coverage etc.

The Cosmic variance can be expressed in terms of an angular integral
over the cube of the two-point correlation function, $C_2(\theta)$
\cite{sred93}. Assuming a gaussian approximation for the two-point
correlation function, we express

\begin{equation}
C_2(\theta) = C_2(0)~\exp[ - {l_c (l_c +1)~\theta^2\over 2}], ~~~~
C_2(0) \approx 3 \times 10^{-5}
\label{c2theta}
\end{equation}

\noindent where the cut-off, $l_c$, in the ${\Delta T/ T}$ angular
spectrum at large values of the spherical harmonic eigenvalue, $l$
($l_c \approx \eta_0/\eta_{\rm rec} \sim 49$ for GW). Using equation
(\ref{c2theta}), the Cosmic variance, $\delta S_3$, for the case of a
CMB anisotropy arising from gravity waves is given by $\delta S_3
\approx 1/(C_2(0)~l_c) \approx 670$. The Cosmic variance for the
scalar case is around $5$ times smaller \cite{mun_sour95}. The Cosmic
variance is larger as the power spectrum tilted away (reddened) from
scale invariance for both scalar perturbations and gravity waves. It
is clear that {\em in principle} the secondary skewness in the CMB for
a CDM model ($\Omega = 1$,$ \Omega_b =0.05$ and $H_0=50 kms/s/Mpc.$)
is {\em unobservable} since it is below the cosmic variance.

\section{Conclusions}

In this work we study a possible source of secondary non-gaussianity
in the CMB and reexamine whether initial gaussian metric perturbations
(as expected from inflation) should lead to a gaussian CMB
anisotropy. We point out that besides nonlinear gravitational
instability, secondary non-gaussianity can be induced in the CMB maps
due to multiple scattering of CMB photon off metric perturbations.  We
develop a general method for calculating CMB temperature fluctuations
beyond the linear order and calculate the expression for CMB
temperature fluctuations at the second order. As an example of
skewness from multiple scattering, we estimate the skewness in the CMB
that is expected to arise due to a relic inflationary gravity wave
background. We show that the magnitude of the effect studied here is
comparable to the ones considered in existing literature. However,
the signal is much smaller than its cosmic variance and we conclude
that gaussian initial perturbation from inflation leads to a gaussian
statistics for the observed temperature fluctuations.

\acknowledgements S.B. wishes to thank Rajaram Nityananda for very
helpful comments and discussions.  D.M. and T.S. are grateful to Varun
Sahni and Alexei Starobinsky for encouragement and useful comments.
The authors also acknowledge  assistance from R. Balasubramanian
in preparing the figure.

\clearpage

\end{document}